\documentclass[12pt]{article}
\usepackage{cite}
\usepackage{epsfig}
\def\delz{\mbox{$<\!\!\delta Z\!\!>$}} 
\def\Journal#1#2#3#4{{#1}{\bf #2} (#4) #3}

\def\EPJ{ Eur.\ Phys.\ J.\ }
\def\NIM{ Nucl.\ Instr.\ Meth.\ }
\def\NP{{ Nucl.\ Phys.\ }}
\def\PL{{ Phys.\ Lett.\ }}

\def\PR{{ Phys.\ Rev.\ }}
\def\ZP{{ Z. Phys.\ }}
\newlength{\numwid}
\newcommand{\nsp}{\hspace*{\numwid}}
\newcommand{\eT}{\mbox{$E_T$}}

\newcommand{\eTg}{\mbox{$E_T^{\,\gamma}$}}

\newcommand{\ZEUSMXD}{ZEUS Collab., M. Derrick et al., }
\newcommand{\ZEUSBre}{ZEUS Collab., J. Breitweg et al., }
\newcommand{\tableht}{\rule[-0.9ex]{0ex}{3.4ex}}
\newcommand{\tablemo}{\rule[-1.4ex]{0ex}{4.1ex}}
\newcommand{\pmerrors}[2]{$^{+#1}_{-#2}$\tableht}

\newcommand{\sleq}{\raisebox{-.4ex}{$\;\stackrel{<}{\scriptstyle \sim}\;$}}

\newcommand{\appr}{\mbox{$\sim$}}
\begin{document}                                                              
\title{\LARGE
\bf Measurement of inclusive prompt photon photoproduction at HERA
\\[10mm]}
\author{ZEUS Collaboration\\[40mm]}
\date{}
\maketitle
\begin{abstract} 
  First inclusive measurements of isolated prompt photons in
  photoproduction at the HERA $ep$ collider have been made with the
  ZEUS detector, using an integrated luminosity of 38.4
  pb$^{-1}$. Cross sections are given as a function of the
  pseudorapidity and the transverse energy ($\eta^\gamma$, \eTg) of
  the photon, for $\eTg > $ 5 GeV in the $\gamma p$ centre-of-mass
  energy range 134--285 GeV.  Comparisons are made with predictions
  from Monte Carlo models having leading-logarithm parton showers, and
  with next-to-leading-order QCD calculations, using currently
  available parameterisations of the photon structure.  For forward
  $\eta^\gamma$ (proton direction) good agreement is found, but in the
  rear direction all predictions fall below the data.
\end{abstract}
\vspace{-17.5cm}
\begin{flushleft}
\tt DESY 99-161\\
Oct.\ 1999\\
\end{flushleft}
\thispagestyle{empty}
\newpage
%
%
%
%
\topmargin-1.cm                                                                                    
\evensidemargin-0.3cm                                                                              
\oddsidemargin-0.3cm                                                                               
\textwidth 16.cm                                                                                   
\textheight 680pt                                                                                  
\parindent0.cm                                                                                     
\parskip0.3cm plus0.05cm minus0.05cm                                                               
\def\3{\ss}                                                                                        
\newcommand{\address}{ }                                                                           
\pagenumbering{Roman}                                                                              
\begin{center}                                                                                     
{                      \Large  The ZEUS Collaboration              }                               
\end{center}                                                                                       
  J.~Breitweg,                                                                                     
  S.~Chekanov,                                                                                     
  M.~Derrick,                                                                                      
  D.~Krakauer,                                                                                     
  S.~Magill,                                                                                       
  B.~Musgrave,                                                                                     
  A.~Pellegrino,                                                                                   
  J.~Repond,                                                                                       
  R.~Stanek,                                                                                       
  R.~Yoshida\\                                                                                     
 {\it Argonne National Laboratory, Argonne, IL, USA}~$^{p}$
% ...
\newpage                                                                                           
$^{\    1}$ now visiting scientist at DESY \\                                                      
% ...
                                                           %
                                                           %
\newpage   
                                                           %
                                                           %
\begin{tabular}[h]{rp{14cm}}                                                                       
$^{a}$ &  supported by the Natural Sciences and Engineering Research                               
          Council of Canada (NSERC)  \\                                                            
$^{b}$ &  supported by the FCAR of Qu\'ebec, Canada  \\                                            
$^{c}$ &  supported by the German Federal Ministry for Education and                               
          Science, Research and Technology (BMBF), under contract                                  
          numbers 057BN19P, 057FR19P, 057HH19P, 057HH29P, 057SI75I \\                              
$^{d}$ &  supported by the MINERVA Gesellschaft f\"ur Forschung GmbH, the                          
German Israeli Foundation, and by the Israel Ministry of Science \\                                
$^{e}$ &  supported by the German-Israeli Foundation, the Israel Science                           
          Foundation, the U.S.-Israel Binational Science Foundation, and by                        
          the Israel Ministry of Science \\                                                        
$^{f}$ &  supported by the Italian National Institute for Nuclear Physics                          
          (INFN) \\                                                                                
$^{g}$ &  supported by the Japanese Ministry of Education, Science and                             
          Culture (the Monbusho) and its grants for Scientific Research \\                         
$^{h}$ &  supported by the Korean Ministry of Education and Korea Science                          
          and Engineering Foundation  \\                                                           
$^{i}$ &  supported by the Netherlands Foundation for Research on                                  
          Matter (FOM) \\                                                                          
$^{j}$ &  supported by the Polish State Committee for Scientific Research,                         
          grant No. 115/E-343/SPUB/P03/154/98, 2P03B03216, 2P03B04616,                             
          2P03B10412, 2P03B03517, and by the German Federal                                        
          Ministry of Education and Science, Research and Technology (BMBF) \\                     
$^{k}$ &  supported by the Polish State Committee for Scientific                                   
          Research (grant No. 2P03B08614 and 2P03B06116) \\                                        
$^{l}$ &  partially supported by the German Federal Ministry for                                   
          Education and Science, Research and Technology (BMBF)  \\                                
$^{m}$ &  supported by the Fund for Fundamental Research of Russian Ministry                       
          for Science and Edu\-cation and by the German Federal Ministry for                       
          Education and Science, Research and Technology (BMBF) \\                                 
$^{n}$ &  supported by the Spanish Ministry of Education                                           
          and Science through funds provided by CICYT \\                                           
$^{o}$ &  supported by the Particle Physics and                                                    
          Astronomy Research Council \\                                                            
$^{p}$ &  supported by the US Department of Energy \\                                              
$^{q}$ &  supported by the US National Science Foundation                                          
\end{tabular}                                                                                      
                                                           %
\newpage
\pagenumbering{arabic}
\setlength{\itemsep}{1ex}
\parsep0ex
\topsep0ex
\newcounter{ccc}
\newenvironment{numlis}{ 
\begin{list}{(\arabic{ccc})}{\usecounter{ccc}\setlength{\itemsep}{1ex}
\setlength{\parsep}{0ex}
\setlength{\topsep}{0ex}}}{\end{list}}
\marginparwidth8pt
\marginparsep1pt 
\newcounter{ccd}
\newcounter{arabiclistc}
\newenvironment{arabiclist}
        {\setcounter{arabiclistc}{0}
         \begin{list}{\arabic{arabiclistc}.}
        {\usecounter{arabiclistc}
         \setlength{\parsep}{0.5mm}
         \setlength{\itemsep}{1mm}}}{\end{list}}
\newenvironment{romlis}{ 
\begin{list}{(\roman{ccd})}{\usecounter{ccd}\setlength{\itemsep}{1ex}
\setlength{\parsep}{0ex}
\setlength{\topsep}{0ex}}}{\end{list}}
%
%
%
\hyphenation{pho-to-pro-duc-tion iso-la-ted pho-ton}
%
\setcounter{page}{1}
\section{Introduction} 

One of the primary aims of photoproduction measurements in $ep$
collisions at HERA is the elucidation of the hadronic behaviour of the
photon.  The measurement of jets at high transverse energy has
provided much information in this area~\cite{Z123,H1}.  In the study
of inclusive jets, next-to-leading order (NLO) QCD calculations are
able to describe the experimental data over a wide range of kinematic
conditions, although the agreement is dependent on the jet
algorithm~\cite{ZEUSincl}.  However, significant discrepancies between
data and NLO theories are found in dijet measurements~\cite{yuji}.  A
further means to study photoproduction is provided by final states
with an isolated high-transverse-energy photon.  These have the
particular merit that the photon may emerge directly from the hard QCD
subprocess (``prompt'' photons), and also can be investigated without
the hadronisation corrections needed in the case of quarks or gluons.
In a previous measurement by ZEUS at HERA~\cite{prph1}, it was shown
that prompt photons, accompanied by balancing jets, are produced at
the expected level in photoproduction and with the expected event
characteristics.  This work is extended in the present paper through
the use of a much larger event sample taken in 1996-97, corresponding
to an integrated $ep$ luminosity of 38.4 pb$^{-1}$.  This allows a
measurement of inclusive prompt photon distributions as a function of
pseudorapidity $\eta^\gamma$ and transverse energy \eTg\ of the
photon, and a comparison with LO and NLO QCD predictions.

 \section{Apparatus and trigger} During 1996-97, HERA collided
positrons with energy $E_e = 27.5$ GeV with protons of energy $E_p
=820$ GeV.  The luminosity was measured by means of the bremsstrahlung
process $ep\to e\gamma p$.

A description of the ZEUS apparatus and luminosity monitor is given
elsewhere~\cite{App}.  Of particular importance in the present work
are the uranium calorimeter (CAL) and the central tracking detector
(CTD).  

The CAL~\cite{UCAL} has an angular coverage of 99.7\% of $4\pi$ and is
divided into three parts (FCAL, BCAL, RCAL), covering the forward
(proton direction), central and rear angular ranges, respectively.
Each part consists of towers longitudinally subdivided into
electromagnetic (EMC) and hadronic (HAC) cells.  The electromagnetic
section of the BCAL (BEMC) consists of cells of $\appr 20$ cm length
azimuthally and mean width 5.45 cm in the $Z$ direction\footnote{ The
ZEUS coordinate system is right-handed with positive-$Z$ in the proton
beam direction and an upward-pointing $Y$ axis. The nominal
interaction point is at $X = Y = Z = 0.$ }, at a mean radius of $\appr
1.3$ m from the beam line.  These cells have a projective geometry as
viewed from the interaction point.  The profile of the electromagnetic
signals observed in clusters of cells in the BEMC provides a partial
discrimination between those originating from photons or positrons,
and those originating from neutral meson decays.

 The CTD~\cite{CTD} is a cylindrical drift chamber situated inside a
superconducting solenoid which produces a 1.43 T field.    Using the
tracking information from the CTD, the vertex of an event can be
reconstructed with a resolution of 0.4 cm in $Z$ and 0.1 cm in $X,Y$.
In this analysis, the CTD tracks are used to reconstruct the event
vertex, and also in the selection criteria for high-$E_T$ photons.

 The ZEUS detector uses a three-level trigger system, of which the
first- and second-level triggers used in this analysis have been
described previously~\cite{prph1}.  The third-level trigger made use
of a standard ZEUS electron finding algorithm~\cite{Ze} to select
events with an electromagnetic cluster of transverse energy $\eT>$ 4
GeV in the BCAL, with no further tracking requirements at this stage.
These events represent the basic sample of prompt photon event
candidates.

\section{Event selection} The offline analysis was based on
previously developed methods~\cite{prph1}.  An algorithm for finding
electromagnetic clusters was applied to the data, and events were
retained for final analysis if a photon candidate with $\eT > 5$ GeV
was found in the BCAL.  A photon candidate was rejected if a CTD
track, as measured at the vertex, pointed to it within 0.3 radians;
this removed almost all high-\eT\ positrons and electrons, including
the majority of those that underwent hard radiation. The BCAL
requirement restricts the photon candidates to the approximate
pseudorapidity\footnote{
All kinematic quantities are given in the laboratory frame.
Pseudorapidity $\eta$ is defined as $-\ln\tan(\theta/2)$, where
$\theta$ is the polar angle relative to the $Z$ direction, measured
from the $Z$ position of the event vertex.} range $-0.75 < \eta^\gamma <
1.0$.

 Events with an identified deep inelastic scattered (DIS) positron in
addition to the BCAL photon candidate were removed, thus restricting
the acceptance to incident photons of virtuality $Q^2\, \sleq1$ GeV$^2$.
The quantity $y^{meas} = \sum(E- p_Z)/2E_e$ was calculated, where the
sum is over all calorimeter cells, \ $E$ is the energy deposited in
the cell, and $p_Z = E\cos\theta$.  When the outgoing positron is not
detected in the CAL, $y^{meas}$ is a measure of $y =
E_{\gamma,\,in}/E_e$, where $E_{\gamma,\,in}$ is the energy of the
incident photon.  If the outgoing positron is detected in the CAL,
$y^{meas} \approx 1$.  A requirement of $0.15 < y^{meas} < 0.7$ was
imposed; the lower cut removed some residual proton-gas backgrounds
while the upper cut removed remaining DIS events, including any with a
photon candidate that was actually a misidentified DIS positron.
Wide-angle Compton scattering events ($ep\to e\gamma p$) were also
excluded by this cut.  This range of accepted $y^{meas}$ values
corresponds approximately to the true $y$ range $0.2 < y < 0.9$.

 An isolation cone was imposed around the photon candidate: within a
cone of unit radius in $(\eta,\,\phi)$, the total \eT\ from other
particles was required not to exceed $0.1\eTg $.  This was calculated
by summing the \eT\ in each calorimeter cell within the isolation
cone. Further contributions were included from charged tracks which
originated within the isolation cone but curved out of it; the small
number of tracks which curved into the isolation cone were ignored.
The isolation condition much reduces the dijet background by removing
a large majority of the events where the photon candidate is closely
associated with a jet and is therefore either hadronic (e.g.\ a
$\pi^0$) or else a photon radiated within a jet.  In particular, the
isolation condition removes most dijet events in which a photon is
radiated from a final-state quark.  
Approximately 6000 events with $\eTg > 5$ GeV remained after the above
cuts.

 Studies based on the single-particle Monte Carlo samples showed that
the photon energy measured in the CAL was on average less than the
true value, owing to dead material in front of the CAL. To compensate
for this, an energy correction, typically 0.2 GeV, was added.

 \section{Monte Carlo simulations}

 In describing the hard interaction of photons of low virtuality with
protons, two major classes of diagram are important.  In one of these
the photon couples in a pointlike way to a $q\bar q$ pair, while in
the other the photon interacts via an intermediate hadronic state,
which provides quarks and gluons which then take part in the hard QCD
subprocesses.  At leading order (LO) in QCD, the pointlike and
hadronic diagrams are distinct and are commonly referred to as direct
and resolved processes, respectively.

In the present analysis, three types of Monte Carlo samples were
employed to simulate: (1) the LO QCD prompt photon processes, (2)
dijet processes in which an outgoing quark radiated a hard photon
(radiative events), and (3) single particles ($\gamma$, $\pi^0$,
$\eta$) at high $E_T$.  All generated events were passed through a
full GEANT-based simulation~\cite{GEANT} of the ZEUS detector.

The PYTHIA 5.7~\cite{Pythia} and HERWIG 5.9~\cite{Herw} Monte Carlo
generators were both used to simulate the direct and resolved prompt
photon processes.  These generators include LO QCD subprocesses and
higher-order processes modelled by initial- and final-state parton
showers.  The parton density function (pdf) sets used were
MRSA~\cite{MRS} for the proton, and GRV(LO)~\cite{GRV} for the
photon. The minimum $p_T$ of the hard scatter was set to 2.5 GeV. No
multi-parton interactions were implemented in the resolved samples.
The radiative event samples were likewise produced using direct and
resolved photoproduction generators within PYTHIA and HERWIG.

 In modelling the overall photoproduction process, the event samples
produced for the separate direct, resolved and radiative processes
were combined in proportion to their total cross sections as
calculated by the generators.  A major difference between PYTHIA and
HERWIG is the smaller radiative contribution in the HERWIG model.

 Three Monte Carlo single-particle data sets were generated,
comprising large samples of $\gamma$, $\pi^0$ and $\eta$. The single
particles were generated uniformly over the acceptance of the BCAL and
with a flat \eT\ distribution between 3 and 20 GeV; \eT-dependent
exponential weighting functions were subsequently applied to reproduce
the observed distributions.  These samples were used in separating the
signal from the background using shower shapes.

 \section{Evaluation of the photon signal} 
Signals in the BEMC that do not arise from charged particles are
predominantly due to photons, $\pi^0$ mesons and $\eta$ mesons. A
large fraction of these mesons decay into multiphoton final states,
the $\pi^0$ through its $2\gamma$ channel and the $\eta$ through its
$2\gamma$ and $3\pi^0$ channels.  For $\pi^0,$ $\eta$ produced with
\eT\ greater than a few GeV, the photons from the decays are separated
in the BEMC by distances comparable to the BEMC cell width in $Z$.
Therefore the discrimination between photons and neutral mesons was
performed on the basis of cluster-shape characteristics, thus avoiding
any need to rely on theoretical modelling of the background.

 A typical high-\eT\ photon candidate consists of signals from a
cluster of 4--5 BEMC cells. Two shape-dependent quantities were used
to distinguish $\gamma$, $\pi^0$ and $\eta$ signals~\cite{prph1}.
These were (i) the mean width \delz\ of the cell cluster in $Z$, which
is the direction of finer segmentation of the BEMC, and (ii) the
fraction $f_{max}$ of the cluster energy found in the most energetic
cell in the cluster.  The quantity \delz\ is defined as $\sum
\left(E_{cell}|Z_{cell} -\overline{Z}|\right)\left/\sum
E_{cell},\right.$ summing over the cells in the cluster, where
$\overline{Z}$ is the energy-weighted mean $Z$ value of the cells.
The \delz\ distribution for the event sample is shown in Figure 1(a),
in which peaks due to the photon and $\pi^0$ contributions are clearly
visible.\footnote{The displacement of the photon peak from the Monte Carlo
prediction does not affect the present analysis; the poor fit in the
region \delz\ = 0.6--1.0 is taken into account in the systematic
errors.}  The Monte Carlo samples of single $\gamma$, $\pi^0$ and
$\eta$ were used to establish a cut on \delz\ at 0.65 BEMC cell
widths, such as to remove most of the $\eta$ mesons but few of the
photons and $\pi^0$s.  Candidates with lower \delz\ were retained,
thus providing a sample that consisted of photons, $\pi^0$ mesons and
a small admixture of $\eta$ mesons.

 The extraction of the photon signal from the mixture of photons and a
neutral meson background was done by means of the $f_{max}$
distributions.
Figure 1(b) shows the shape of the $f_{max}$ distribution for the final
event sample, after the \delz\ cut, 
fitted to the  $\eta$ component determined from the \delz\
distribution plus freely-varying $\gamma$ and $\pi^0$ contributions.
Above an $f_{max}$ value of 0.75, the distribution is dominated by the
photons; below this value it consists mainly of meson background.
Since the shape of the $f_{max}$ distribution is similar for the
$\eta$ and $\pi^0$ contributions, the background subtraction is
insensitive to uncertainties in the fitted $\pi^0$ to $\eta$ ratio.

 The numbers of candidates with $f_{max} > 0.75$ and $f_{max} < 0.75$
were calculated for the sample of events occurring in each bin of a
measured quantity.  From these numbers, and the ratios of the
corresponding numbers for the $f_{max}$ distributions of the single
particle samples, the number of photon events in the given bin was
evaluated~\cite{prph1}.

\section{Cross section calculation and systematic uncertainties}
Cross sections are given for the photoproduction process $ep\to
\gamma(\mbox{\rm prompt}) + X,$ taking place in the incident $\gamma
p$ centre-of-mass energy $(W)$ range 134--285 GeV, i.e.\ $0.2 < y <
0.9 $.  The virtuality of the incident photon is restricted to the
range $Q^2 \sleq 1$ GeV$^2$, with a median value of approximately
$10^{-3}$ GeV$^2$.  The cross sections represent numbers of events
within a given bin, divided by the bin width and integrated luminosity.
They are given at the hadron level, with an isolation cone defined 
around the prompt photon as at the detector level.  To obtain the
hadron-level cross sections, bin-by-bin correction factors were
applied to the corresponding detector-level distributions; these
factors were calculated using PYTHIA.

 The following sources of systematic error were taken into account: 
\begin{arabiclist}
\item {\it Calorimeter simulation:} the uncertainty of the simulation
of the calorimeter response~\cite{ZCALS} gives rise to an uncertainty
on the cross sections of $\pm 7\%$; 
\item {\it Modelling of the shower shape:} uncertainties on the
agreement of the simulated $f_{max}$ distributions with the data
correspond to a systematic error averaging $\pm 8\%$ on the final
cross sections;
\item {\it Kinematic cuts:} the cuts defining the accepted kinematic
range at the detector level were varied by amounts corresponding to
the resolution on the variables.  Changes of up to 5\% in the cross
section were observed;
\item {\it $\eta/(\eta + \pi^0)$ ratio:} the fitted value was
typically 25\%; variations of this ratio in the range 15--35\% led to
cross section variations of around $\pm2\%$;
\item {\it Vertex cuts:} narrowing the vertex cuts to ($-25, +15$) cm
from their standard values of ($-50, +40$) cm gave changes in the
cross sections of typically $\pm 4\%$.  
\end{arabiclist} 
In addition, studies were made of the effects of using HERWIG instead
of PYTHIA for the correction factors, of varying the \eT\ distribution
applied to the single-particle samples, and of varying the composition of
the Monte Carlo simulation in terms of direct, resolved and radiative
processes.  These gave changes in the cross sections at the 1\% level.
The 1.6\% uncertainty on the integrated luminosity was neglected.  The
individual contributions were combined in quadrature to give the total
systematic error.

\section{Theoretical calculations} 
In presenting cross sections, comparison is made with two types of
theoretical calculation, in which the pdf sets taken for both the
photon and proton can be varied, although there is little sensitivity
to the choice of proton pdf.  These are:

 {\it (i)\/} PYTHIA and HERWIG calculations evaluated at the
 final-state hadron level, as outlined in Sect.\ 4.  Each of these
 programs comprises a set of LO matrix elements augmented by
 parton showers  in the initial and final states together with
 hadronisation;

 {\it (ii)\/} NLO parton-level calculations of Gordon (LG)~\cite{LG,GV}
 and of Krawczyk and Zembrzuski (K\&Z)~\cite{KZ}.  Pointlike and
 hadronic diagrams at the Born level are included, together with
 virtual (loop) corrections and terms taking into account three-body
 final states.  The radiative terms are evaluated by means of
 fragmentation functions obtained from experiment.  In both
 calculations, the isolation criterion was applied at the parton level.

 The LG and K\&Z calculations differ in several
respects~\cite{LG,GV,KZ,Klasen}.  The K\&Z calculation includes a
box-diagram contribution for the process $\gamma g \to \gamma
g$~\cite{Combridge}, but excludes higher-order corrections to the
resolved terms which are present in LG.  A value of
$\Lambda_{\overline{MS}} = 200$ MeV (5 flavours) is used in LG while
in K\&Z a value of 320 MeV (4 flavours) is
used, so as to reproduce a fixed value of $\alpha_S = 0.118$ at the
$Z^0$ mass.  The standard versions of both calculations use a QCD
scale of $p_T^2$.  Both calculations use higher-order (HO) versions of
the GRV~\cite{GRV} and GS~\cite{GS} photon pdf sets.

\section{Results} 

    Figure 2 and Table 1 give the inclusive cross-section
$d\sigma/d\eTg$ for the production of isolated prompt photons in the
range $-0.7 < \eta^\gamma < 0.9$ for $0.2 < y < 0.9$.  All the
theoretical models describe the shape of the data well; however the
predictions of PYTHIA and especially HERWIG are too low in magnitude.
The LG and K\&Z calculations give better agreement with the data.

 Figure 3 and Table 2 give the inclusive cross-section
$d\sigma/d\eta^\gamma$ for isolated prompt photons in the range $5 <
\eTg < 10$ GeV for $0.2 < y < 0.9$.  Using the GRV pdf's in the photon,
PYTHIA gives a good description of the data for forward
pseudorapidities.  The HERWIG distribution, while similar in shape to
that of PYTHIA, is lower throughout; this is attributable chiefly to
the lower value of the radiative contribution in HERWIG (see Sect.\
4).  The LG and K\&Z calculations using GRV are similar to each other
and to PYTHIA.  All the calculations lie below the data in the lower
$\eta^\gamma$ range.

 The effects were investigated of varying some of the parameters of
the K\&Z calculation relative to their standard values (NLO, 4
flavours, $\Lambda_{\overline{MS}} = 320$ MeV, GRV photon pdf).
Reducing the number of flavours used in the calculation to three (with
$\Lambda_{\overline{MS}} = 365$ MeV) reduced the cross sections by
35--40\% across the $\eta^\gamma$ range, confirming the need to take
charm into account.  A LO calculation (with $\Lambda_{\overline{MS}}
= 120$ MeV and a NLO radiative contribution) was approximately 25\%
lower than the standard NLO calculation.  Variations of the QCD scale
between $0.25E_T^2$ and $4E_T^2$ gave cross-section variations of
approximately $\pm3\%$.

 Figure 3(b) illustrates the effects of varying the photon parton
densities, comparing the results using GRV with those using GS.  The
ACFGP parton set~\cite{ACFGP} gives results (not shown) similar to
GRV.  All NLO calculations describe the data well for $\eta^\gamma >
0.1$, as does PYTHIA, but are low at more negative $\eta^\gamma$
values, where the curves using the GS parton densities give poorer
agreement than those using GRV.

 As a check on the above results, the same cross sections were
evaluated with the additional requirement that each event should
contain a jet (see \cite{prph1}) with $E_T \ge 5$ GeV in the
pseudorapidity range ($-1.5$, 1.8).  Both the measured and theoretical
distributions were found to be of a similar shape to those in Fig.\ 3.

 The discrepancy between data and theory at negative $\eta^{\gamma}$
is found to be relatively strongest at low values of $y$.  Figure 4
shows the inclusive cross section $d\sigma/d\eta^\gamma$ as in Fig.\
3, evaluated for the three $y$ ranges 0.2--0.32, 0.32--0.5 and
0.5--0.9 by selecting the $y^{meas}$ ranges 0.15--0.25, 0.25--0.4 and
0.4--0.7 at the detector level.  The numerical values are listed in
Table 2.  In the lowest $y$ range, both theory and data show a peaking
at negative $\eta^\gamma$, but it is stronger in the data.  The Monte
Carlo calculations indicate that the peak occurs at more negative
$\eta^\gamma$ values as $y$ increases, eventually leaving the
measurement acceptance.  In the highest $y$ range (Fig.\ 4(c)), 
agreement is found between theory and data.  The movement of the
peak can be qualitatively understood by noting that for fixed values
of \eT\ and $x_\gamma$, where $x_\gamma$ is the fraction of the
incident photon energy that contributes to the resolved QCD
subprocesses, measurements at increasing $y$ correspond on average to
decreasing values of pseudorapidity.  By varying the theoretical
parameters, the discrepancy was found to correspond in the K\&Z
calculation to insufficient high $x_\gamma$ partons in the resolved
photon.

\section{Summary and conclusions}

 The photoproduction of isolated prompt photons within the kinematic
range $0.2 < y < 0.9$, equivalent to incident $\gamma p$
centre-of-mass energies $W$ of 134--285 GeV, has been measured in the
ZEUS detector at HERA, using an integrated luminosity of 38.4
pb$^{-1}$.  Inclusive cross sections for $ep\to \gamma + X$ have been 
presented as a function of \eTg\ for photons in the pseudorapidity
range $-0.7 < \eta^\gamma < 0.9$, and as a function of $\eta^\gamma$
for photons with $5 < \eTg < 10$ GeV.  The latter results have been given
also for three subdivisions of the $y$ range.  All kinematic quantities
are quoted in the laboratory frame. 

 Comparisons have been made with predictions from leading-logarithm
parton-shower Monte Carlos (PYTHIA and HERWIG), and from
next-to-leading-order parton-level calculations.  The models are able
to describe the data well for forward (proton direction) photon
pseudorapidities, but are low in the rear direction.  None of the
available variations of the model parameters was found to be capable
of removing the discrepancy with the data.  The disagreement is
strongest in the $W$ interval 134--170 GeV, but not seen within
the measurement acceptance for $W > 212$ GeV. This result, together with the
disagreements with NLO predictions seen also in recent dijet
results at HERA\cite{yuji}, would appear to indicate a need to review
the present theoretical modelling of the parton structure of the photon.
 
\section*{Acknowledgements}
 It a pleasure to thank the DESY
 directorate and staff for their unfailing support and encouragement.
 The outstanding efforts of the HERA machine group in providing high
 luminosity for the 1996--97 running are much appreciated.  We are also
 extremely grateful to L. E. Gordon, M. Krawczyk and A. Zembrzuski for
 helpful conversations, and for making available to us calculations of
 the NLO prompt photon cross section and a computer program (K\&Z).

\newpage

 \begin{table}\begin{center}
 \begin{tabular}{|c|r|}\hline   
       \eTg &\tablemo  $d\sigma / d\eTg$ pb GeV$^{-1}$ \\  
 \hline
  5.0 -  6.0 &  18.4 $\pm$  2.1 \pmerrors{ 2.6}{ 2.8} \\
  6.0 -  7.0 &  9.9 $\pm$  1.3 \pmerrors{ 1.2}{ 1.3} \\
  7.0 -  8.0 &  8.7 $\pm$  1.1 \pmerrors{ 0.9}{ 0.7} \\
  8.0 -  9.5 &  3.3 $\pm$  0.6 \pmerrors{ 0.4}{ 0.5} \\
 \nsp9.5 - 11.0 &  2.2 $\pm$  0.4 \pmerrors{ 0.2}{ 0.3} \\
 11.0 - 13.0 &  1.3 $\pm$  0.3 \pmerrors{ 0.2}{ 0.2} \\
 13.0 - 15.0 &  0.3 $\pm$  0.3 \pmerrors{ 0.2}{ 0.1} \\
 \hline
 \end{tabular}

 \end{center} \caption{Differential cross sections for inclusive
 photoproduction of isolated photons with $-0.7 < \eta^\gamma < 0.9$,
 averaged over given transverse-energy intervals, for $0.2 < y < 0.9$
 ($134 < W < 285$~GeV).  The first error is statistical, the second is
 systematic.  \protect\\[00mm] } \end{table}

\begin{table}\begin{center}
 \begin{tabular}{|c|r|r|r|r|}\hline
 \multicolumn{1}{|c}{\tablemo$\eta^{\gamma}$} &
 \multicolumn{4}{|c|}{$d\sigma / d\eta^\gamma$ pb} \\ \cline{2-5} &
 \multicolumn{1}{|c}{ $0.2 < y < 0.9$} & \multicolumn{1}{|c}{ ~$0.2 <
 y < 0.32$} & \multicolumn{1}{|c}{ $0.32 < y < 0.5$ ~} &
 \multicolumn{1}{|c|}{ $0.5 < y < 0.9$} \tableht\\ &
 \multicolumn{1}{|c}{ $134 < W < 285$} & \multicolumn{1}{|c}{ $134 < W
 < 170 $} & \multicolumn{1}{|c}{ $170 < W < 212$} &
 \multicolumn{1}{|c|}{ $212 < W < 285$} \tableht\\ \hline 

$-0.6$ & 38.9 $\pm$ 5.9 \pmerrors{ 4.3}{ 4.7} & 10.0 $\pm$ 3.3
   \pmerrors{ 1.8}{ 2.2} & 15.2 $\pm$ 3.6 \pmerrors{ 2.4}{ 2.5} & 13.7
   $\pm$ 3.2 \pmerrors{ 2.3}{ 3.1} \\

$-0.4$ & 40.1 $\pm$ 5.7 \pmerrors{ 4.3}{ 5.1} & 17.0 $\pm$ 3.6
   \pmerrors{ 3.1}{ 1.9} & 13.8 $\pm$ 3.2 \pmerrors{ 2.1}{ 2.7} & 9.3
   $\pm$ 2.8 \pmerrors{ 1.5}{ 2.2} \\

$-0.2$ & 27.7 $\pm$ 5.0 \pmerrors{ 2.8}{ 4.1} & 11.4 $\pm$ 3.1
    \pmerrors{ 2.1}{ 2.2} & 11.7 $\pm$ 2.9 \pmerrors{ 1.9}{ 2.0} & 4.6
    $\pm$ 2.1 \pmerrors{ 0.7}{ 1.3} \\

\nsp0.0 & 35.1 $\pm$ 5.5 \pmerrors{ 4.1}{ 3.6} & 17.7 $\pm$ 3.5
    \pmerrors{ 3.1}{ 2.2} & 9.1 $\pm$ 3.0 \pmerrors{ 1.7}{ 1.6} & 8.3
    $\pm$ 2.7 \pmerrors{ 1.3}{ 1.9} \\

\nsp0.2 & 21.0 $\pm$ 4.4 \pmerrors{ 3.3}{ 3.0} & 9.9 $\pm$ 2.5
  \pmerrors{ 2.1}{ 1.8} & 5.7 $\pm$ 2.4 \pmerrors{ 1.2}{ 1.1} & 5.4
  $\pm$ 2.4 \pmerrors{1.1}{ 1.4} \\

\nsp0.4 & 18.7 $\pm$ 3.7 \pmerrors{ 2.1}{ 2.2} & 7.2 $\pm$ 2.1
  \pmerrors{ 1.4}{ 0.9} & 7.9 $\pm$ 2.4 \pmerrors{ 1.4}{ 1.6} & 3.6
  $\pm$ 1.7 \pmerrors{ 0.6}{ 0.9} \\

\nsp0.6 & 14.4 $\pm$ 4.4 \pmerrors{ 2.1}{ 2.3} & 6.8 $\pm$ 2.7
   \pmerrors{ 1.4}{ 1.0} & 5.8 $\pm$ 2.5 \pmerrors{ 0.9}{ 1.3} & 1.8
   $\pm$ 2.2 \pmerrors{ 0.6}{ 0.7} \\

\nsp0.8 & 19.5 $\pm$ 4.9 \pmerrors{ 2.1}{ 2.6} & 6.8 $\pm$ 2.9
  \pmerrors{ 1.3}{ 1.3} & 10.1 $\pm$ 3.4 \pmerrors{ 2.3}{ 1.5} & 2.6
  $\pm$ 2.2 \pmerrors{ 0.5}{ 0.9} \\

 \hline

\end{tabular}\end{center}
\caption{Differential cross sections per unit pseudorapidity for
inclusive photoproduction of isolated photons with $5 < \eTg < 10$
GeV, averaged over laboratory pseudorapidity intervals
of $\pm0.1$ about the given central values.  The
$\gamma p$ centre-of-mass energy $(W)$ ranges are in GeV.  Results are
listed for the full range of fractional incident photon energy $y$ and in 
three subdivisions.  The first error is statistical, the second is
systematic.  \protect\\[1mm]}
\end{table}

\begin{figure}\centerline{
\epsfig{file=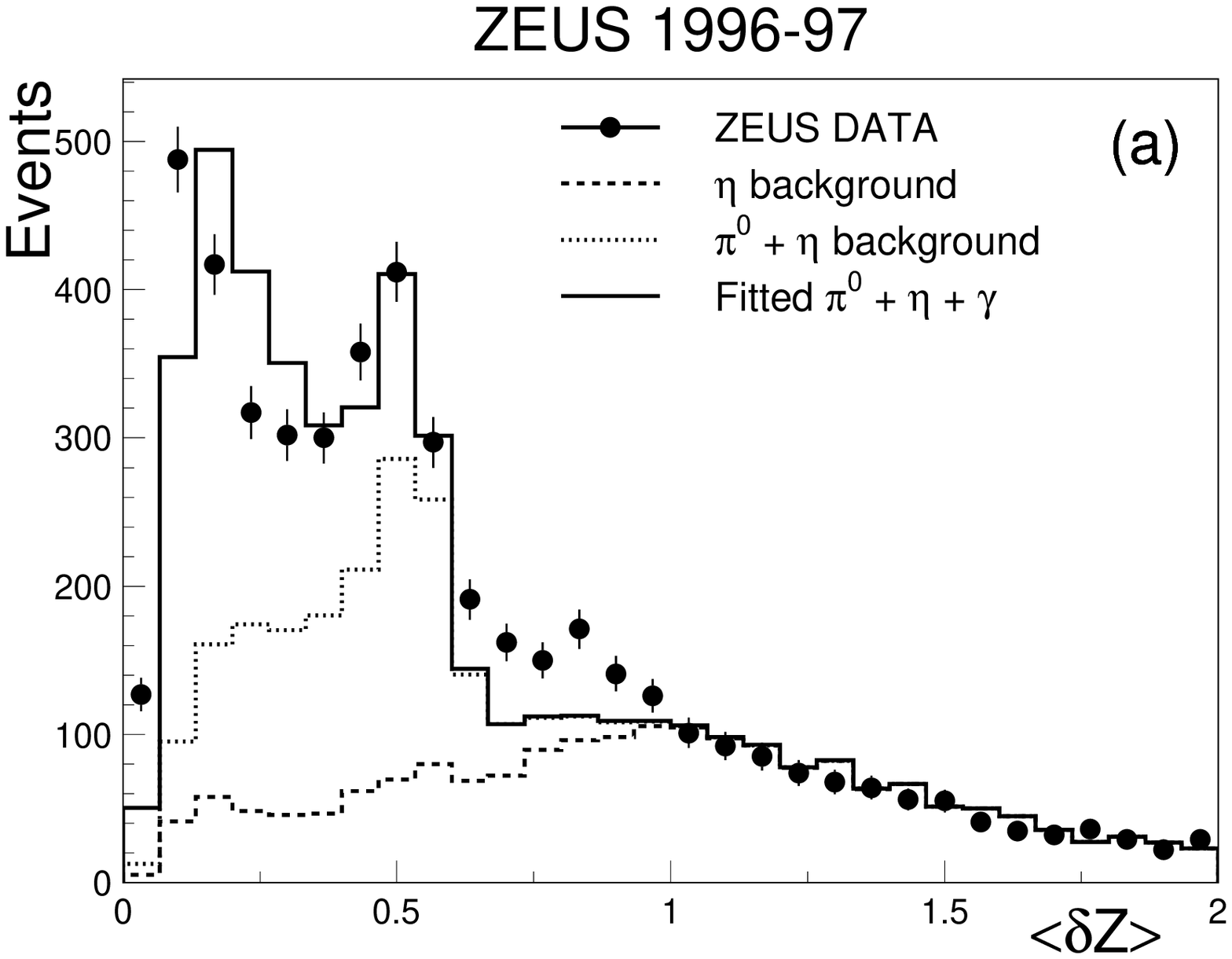,height=120mm%
}}\vspace{-36mm}\centerline{
\epsfig{file=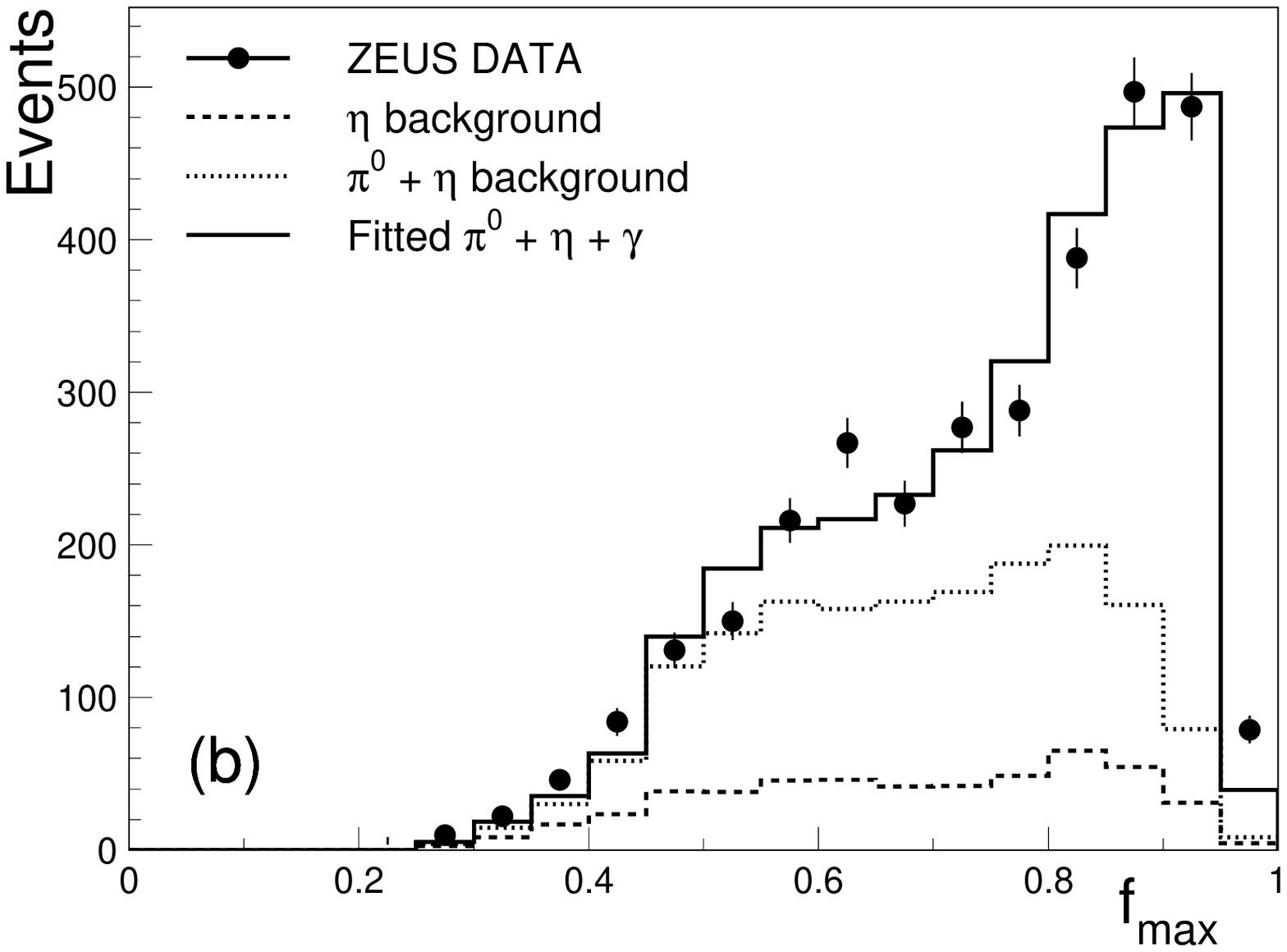,height=120mm%
}}              
\vspace*{-1.5cm}

\caption{\small (a) Distribution of \delz\ for prompt photon
candidates in selected events.  (b) Distribution of $f_{max}$ for
prompt photon candidates in selected events after cutting on \delz.
Also given in both cases are fitted Monte Carlo distributions for photons,
$\pi^0$ and $\eta$ mesons with similar selection requirements as for the
observed photon candidates.  Samples with $f_{max} > 0.75$ and
$f_{max} < 0.75$ are enriched in the photon signal and in the meson
background, respectively.  }
\label{fmax}\end{figure}

\begin{figure}\centerline{
\epsfig{file=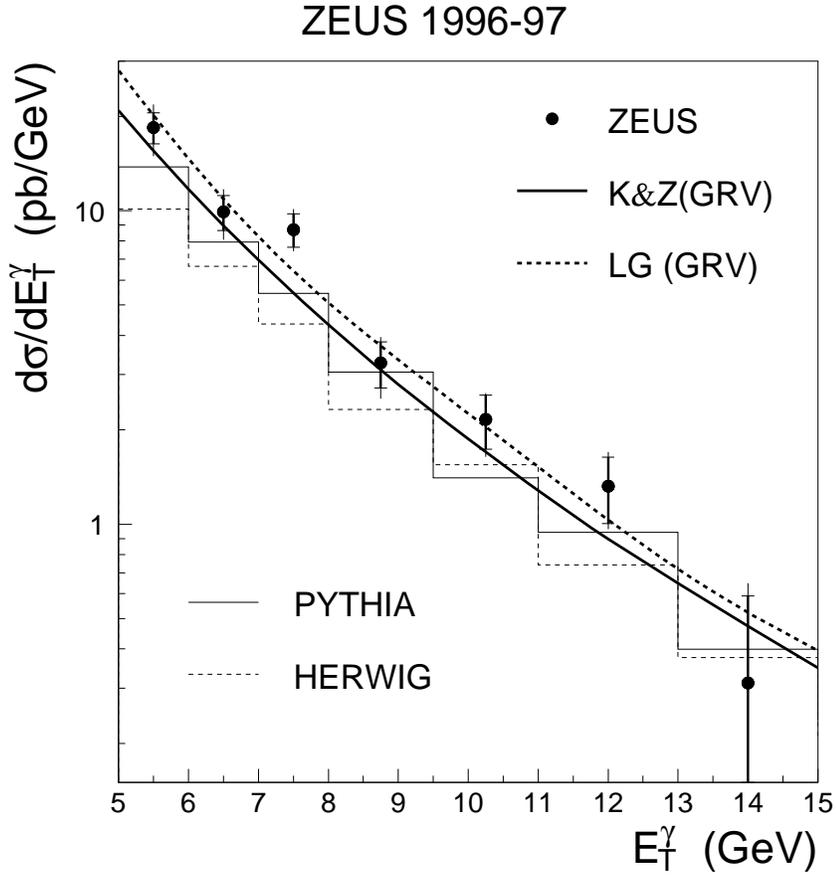,height=12cm,%
}}
\caption{\small Differential cross section $d\sigma/d\eTg$ for prompt
photons produced over $-0.7 < \eta^\gamma < 0.9$. The inner (thick)
error bars are statistical; the outer include systematic errors added
in quadrature.  The data points are plotted at the respective bin
centres (see Table 1; bin centring corrections are negligible).
Predictions are shown from PYTHIA and HERWIG at the hadron level
(histograms), and from LG and K\&Z (curves).  In K\&Z, the default
4-flavour NLO $\Lambda_{\overline{MS}}$ value of 320 MeV is used.
}\end{figure}

\begin{figure}\centerline{
\epsfig{file=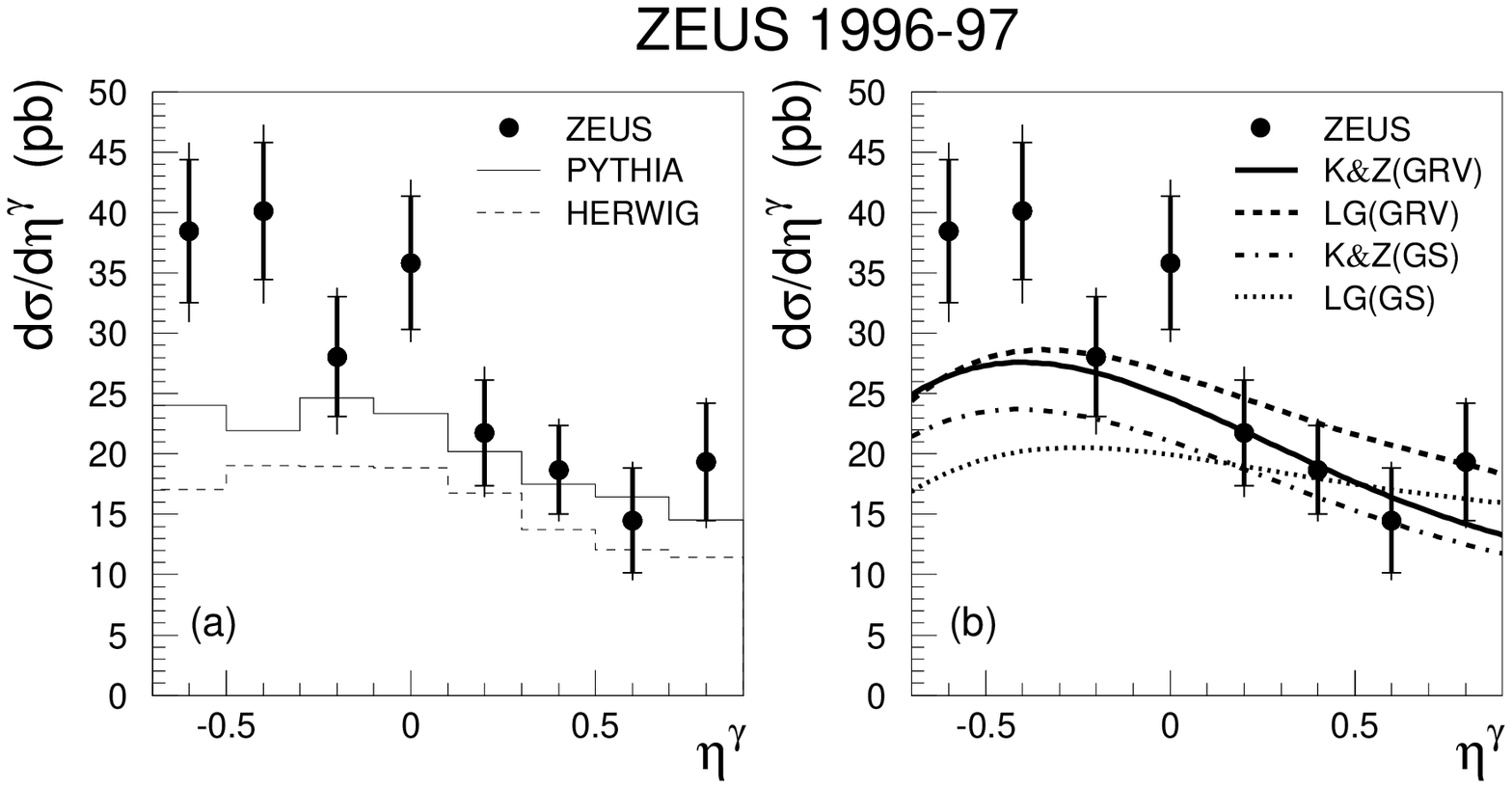,height=16cm}}
\caption{\small Differential cross-section $d\sigma/d\eta^\gamma$ for
isolated prompt photons with $5 < \eTg < 10 $ GeV, for $0.2 < y< 0.9$
($134 < W < 285$ GeV).  The inner (thick) error bars are statistical; 
the outer include systematic errors added in quadrature.  Also plotted are
(a) PYTHIA and HERWIG predictions using the GRV(LO) photon parton
densities; (b) LG and K\&Z NLO  predictions using GRV(HO) and GS(HO)
photon parton densities. }\end{figure}

\begin{figure}\centerline{
\epsfig{file=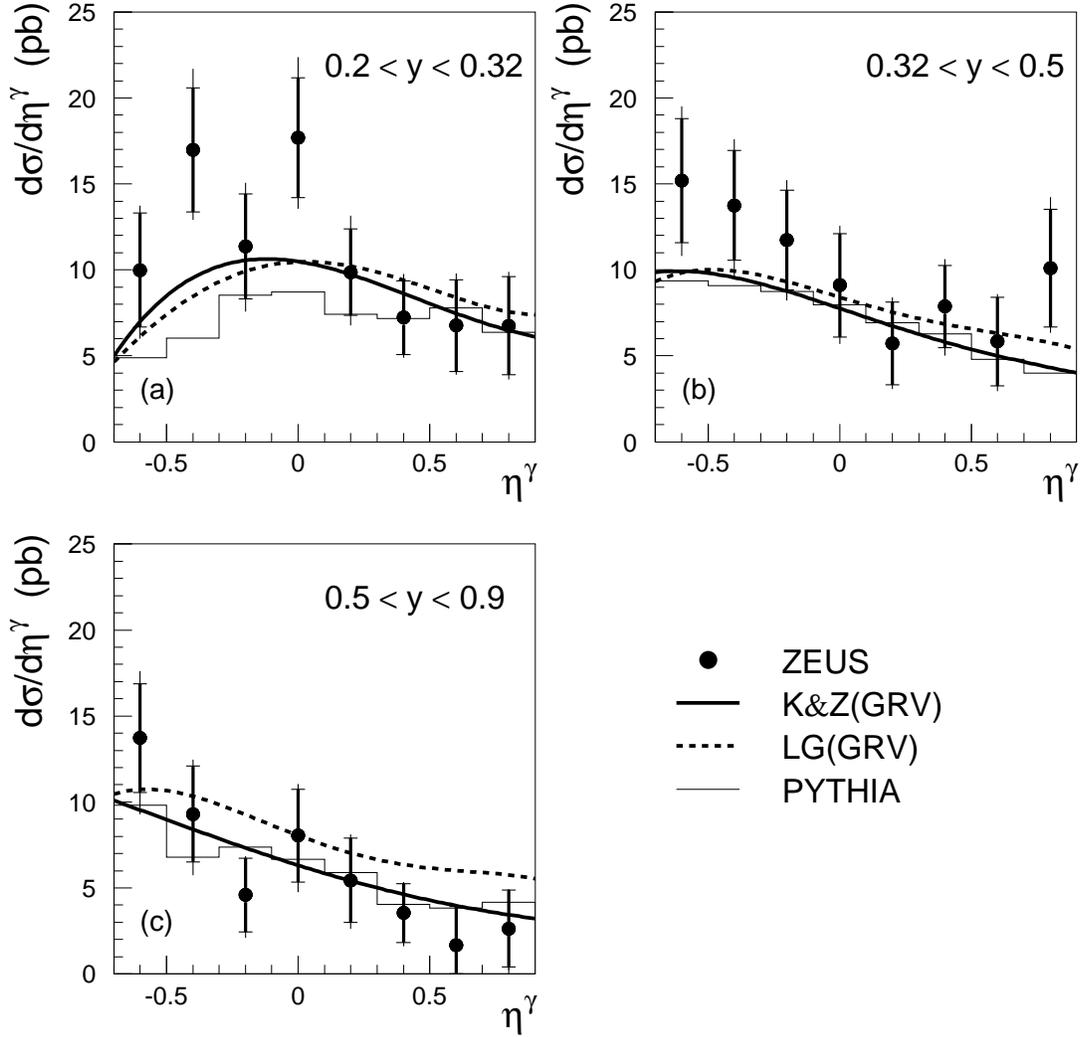,height=16cm,%
}}
\caption{\small Differential cross-section $d\sigma/d\eta^\gamma$, for
isolated prompt photons with $5 < \eTg < 10 $ GeV, compared with
PYTHIA and with  LG and K\&Z NLO predictions, using GRV photon parton densities
as in Fig.\ 3. The inner (thick) error bars are statistical; the outer
include systematic errors added in quadrature.  The plots correspond to the
$W$ ranges (a) 134--170~GeV, (b) 170--212~GeV, (c) 212--285~GeV.}
\end{figure}


\newpage

 \begin{thebibliography}{29} 

 \bibitem{Z123}\ZEUSMXD \Journal{\PL}{B322}{287}{1994};\\
	       \ZEUSMXD \Journal{\PL}{B348}{665}{1995};\\
	       \ZEUSMXD \Journal{\PL}{B384}{401}{1996}.
 \bibitem{H1} 
 H1 Collab., T. Ahmed et al., \Journal{\PL}{B297}{205}{1992};\\
 H1 Collab., T. Abt et al., \Journal{\PL}{B314}{436}{1993};\\
 H1 Collab., T. Ahmed et al., \Journal{\NP}{B445}{195}{1995};\\
 H1 Collab., S. Aid et al., \Journal{\PL}{B392}{234}{1997}.
 \bibitem{ZEUSincl} \ZEUSBre \Journal{\EPJ}{C4}{591}{1998}. 
 \bibitem{yuji} 
 \ZEUSBre \Journal{\EPJ}{C1}{109}{1998};\\
 \ZEUSBre DESY 99-057, to appear in Eur.\ Phys.\ J.
 \bibitem{prph1}\ZEUSBre \Journal{\PL}{B413}{201}{1997}.
 \bibitem{App} \ZEUSMXD \Journal{\PL}{B297}{404}{1992}.
 \bibitem{UCAL}  A. Andresen et al., \Journal{\NIM}{A309}{101}{1991};\\
 A. Bernstein et al., \Journal{\NIM}{A338}{23}{1993};\\
 A. Caldwell et al., \Journal{\NIM}{A321}{356}{1992}.
 \bibitem{CTD} N. Harnew et al., \Journal{\NIM}{A279}{290}{1989};\\
  B. Foster et al., \Journal{\NP}{B\mbox{ \rm(Proc.\ Suppl.) }32}{181}{1993};  
 \Journal{\NIM}{A338}{254}{1994}.
 \bibitem{Ze}\ZEUSMXD \Journal{\ZP}{C65}{379}{1995}. 
 \bibitem{GEANT} R. Brun et al., CERN-DD/EE/84-1 (1987).
 \bibitem{Pythia} H.-U. Bengtsson and T. Sj\"ostrand, Comp.\ Phys.\ Commun.\ 
 {\bf 46} (1987) 43;\\   T. Sj\"ostrand, CERN-TH.6488/92.
 \bibitem{Herw}G. Marchesini et al., Comp.\ Phys.\ Commun. {\bf 67} (1992) 465.
 \bibitem{MRS} A. D. Martin, R. G. Roberts and W. J. Stirling, 
 \Journal{\PR}{D50}{6734}{1994}.
 \bibitem{GRV} M. Gl\"uck, E. Reya and A. Vogt, \Journal{\PR}{D46}{1973}{1992}.
 \bibitem{ZCALS} \ZEUSBre \Journal{\EPJ}{C6}{67}{1999};\\
 \ZEUSMXD \Journal{\ZP}{C72}{399}{1996}.
 \bibitem{LG} L. E. Gordon, \Journal{\PR}{D57}{235}{1998} and private
 communication;\\ L. E. Gordon, {\tt hep-ph/9706355} and Proceedings,
 {\it Photon 97}, Egmond aan Zee, eds.\ A. Buijs and F. Ern\'e (World
 Scientific, Singapore, 1998), 173.
 \bibitem{GV} L. E. Gordon and W. Vogelsang, {\tt hep-ph/9606457} and
 Proceedings, {\it DIS96}, Rome, eds.\ G. D'Agostini and A. Nigro,
 (World Scientific, Singapore, 1997), 278.
 \bibitem{KZ} M. Krawczyk and A. Zembrzuski, {\tt hep-ph/9810253} and
 Proceedings, {\it Photon 97}, Egmond aan Zee, eds.\ A. Buijs and F. Ern\'e 
 (World Scientific, Singapore, 1998), 162 and private communication. 
 \bibitem{Klasen} M. Klasen, {\tt hep-ph/9907366.}
 \bibitem{Combridge} B. L. Combridge, \Journal{\NP}{B174}{243}{1980}.
 \bibitem{GS} L. E. Gordon and J. K. Storrow, \Journal{\NP}{B489}{405}{1997}. 
 \bibitem{ACFGP} P. Aurenche et al., LPTHE preprint 92/13;\\  
 P. Aurenche et al., \Journal{\ZP}{C64}{621}{1994}.
 \end{thebibliography}
\end{document}